Albert J. Ammerman
November 1, 2020: arXiv text used as starting point for changes
**Revised August 4, 2021** [Using in part changes suggested on May 1, 2021]

# The Neolithic Transition in Europe at 50 Years

## 1. Introduction

One of the last chapters in the long course of human evolution was the shift from hunting and gathering to the production of food or strategies of subsistence based on farming and the herding of animals. In Southwest Asia, the first steps towards the origins of agriculture began some 12,000 years ago and then spread over most regions of Europe during the span of time from about 10,400 years ago (the start of the so-called PPNB on the island of Cyprus) through around 6,000 years ago. The aim of this chapter is to provide an overview on the research that we have done on the question of the Neolithic transition in Europe, which began when Luca Cavalli-Sforza, a leading figure in the field of human population genetics, and I began to work in collaboration at the University of Pavia in November of 1970. In a matter of a few days, we sketched out what would amount to taking a completely new approach to the problem (Ammerman 2003). Below more will be said about our first attempt at measuring the rate of spread of early farming in Europe (Ammerman and Cavalli-Sforza 1971) and the proposal of the working hypothesis of demic diffusion and the new wave of advance model to explain how the spread over Europe took place (Ammerman and Cavalli-Sforza 1973). This turned out to be a highly productive collaboration between two scholars coming from quite different backgrounds: one in prehistoric archaeology and environmental studies and the other in genetics and the medical sciences. In short, our work provides a good example of interdisciplinary research in the 1970s and 1980s, when what we were setting out to do initially seemed to make little or no sense to many of our colleagues in the two fields of study. Today the need for such collaboration between human genetics and archaeology is, of course, taken for granted.

When Cavalli-Sforza now took up a new position in the Department of Genetics at Stanford University, I was finishing my PhD at the Institute of Archaeology in London (with a dissertation on the Mesolithic period in Italy) and joined him in January of 1972 and began teaching in the Program in Human Biology at Stanford. It was a new interdisciplinary program for undergraduates, and I was officially based in the Department of Genetics – the only archaeologist with such a position for many years to come.  On the positive side, I did have a background in mathematics and statistics; for several years, I had worked with Roy Hodson at the Institute of Archaeology, one of the pioneers in applying quantitative methods to archaeology.  Indeed, Luca and I first met at the conference on quantitative archaeology jointly sponsored by the Royal Society and the Academy of Romania, which was held at Mamaia on the Black Sea in the summer of 1970 (Hodson et al., 1971).  More will be said below about the multiple lines of study that we carried out over the next ten years and brought together in a synthesis called *The Neolithic Transition and the Genetics of Populations in Europe* (Ammerman and Cavalli-Sforza 1984).  It is worth adding here that we now called this field of study "the Neolithic transition" and no longer "the Neolithic revolution" of V. Gordon Childe, since it unfolded so slowly on the ground, taking well over 120 human generations to move from Anatolia to Denmark.  While there were obviously major changes in subsistence, material culture, settlement patterns and so forth, the slow pace at which it moved across Europe as a whole did not warrant the use of the term "revolution" but something else.

By way of introduction, it is time to skip forward to the Wenner Gren Workshop called "The Neolithic Transition in Europe:  Looking back  -- Looking Forward," which was held at Venice in October of 1998.  It brought together a group of 21 leading scholars with a wide range of experience working on the question.  The proceedings of the meeting were then published in a volume called *The Widen Harvest* (Ammerman and Biagi 2003) – with its subtitle taken from the name of the Wenner Gren Workshop itself.  The aim of the conference was to do more than just take stock of the current state of knowledge in a rapidly developing field of study.  The participants were encouraged to step back and reflect upon their thinking on the question in a broader sense and how their own research had evolved over the years.  In short, a conscious effort was made to place greater emphasis on the historical context in which we, as archaeologists, attempt to address and solve problems connected with the Neolithic transition.  In the rush to obtain answers from the archaeological evidence that is immediately at hand, there is a tendency to lose sight of the context in which we make choices with regard to such things as how and where we conduct fieldwork.  In other words, not enough attention is paid to the wider historical framework in which our work is taking place (Ammerman 2003: 4).

The idea for the Venice meeting first came to mind in 1996 when I was a Kress Senior Fellow at the Center for Advanced Studies in Visual Arts (located in the National Gallery of Art in Washington, D.C.). At the time, I had just spent the last year studying the many different attempts at reconstructing the ancient city of Rome over the course of the last five centuries.  It is worth adding here that, in the years between 1985 and 2002, I had the opportunity to do something quite different than Neolithic studies: namely, to conduct environmental studies in the centers of three cities of historical importance. In 1985, when I was the head of a research group at the Institute of Ecology of the University of Parma (Italy), I was invited by the Italian government to coordinate the environmental studies to be conducted in conjunction with the excavations being done at several early sites in and around the Forum of ancient Rome. In turn, without going into the details here, the work in Rome led to the invitation to carry out fieldwork of much the same kind in the Agora of ancient Athens and then at Saint Mark's Square in Venice. Thus, in 1993, I would have the rare opportunity to conduct fieldwork in the Forum of ancient Rome, the Agora of ancient Athens and Piazza San Marco in Venice all in the same year (for the comparative study of what we learned in the three cases and bibliography, see Ammerman 2011a). Returning to Rome, the reconstruction of the ancient city has taken many different forms over the centuries: from paintings to maps, from ice sculptures and scale models in cork to the restoration of the monuments themselves (Ammerman 2003: 4). One of the lessons that I learned from tracing such a long and rich series of reconstructions was the importance of paying closer attention to the historiography of an archaeological problem. In looking back on the study of the Neolithic transition in Europe now over a span of 50 years, there is the need to remember this lesson.

In practical terms, the Venice meeting was held at the time of the 25[th] anniversary of the publication of "A population model for the diffusion of early farming in Europe" (Ammerman and Cavalli-Sforza 1973). The paper was actually given in December 1971 at the conference on "The Explanation of Culture Change: Models in Prehistory," which was organized by Colin Renfrew. In 1998, Cavalli-Sforza was now 76 years old, and his health was starting to waver, so the meeting was held in his honor. Colin Renfrew paid tribute to him as a friend and scientist with wide interests at the final dinner of the Wenner Gren Workshop. Today *The Widening Harvest* provides a guide to where research on the question stood at the end of the 20[th] century. It is worth recalling that when our book came out in 1984, there was still no direct line of access to the genes of the first farmers themselves. Instead, one had to make inferences on the basis of the statistical analysis of patterns in the genes of modern populations in Europe, which were used as proxies for the remote past (e.g., Menozzi et al., 1978; Cavalli-Sforza et al., 1994). At the Venice meeting in 1998, Bryan Sykes (2003), a geneticist at Oxford University, presented a paper on what he was starting to find from the direct analysis of human bones of Neolithic age. In short, work of this kind was now in progress at several laboratories but it was still too soon to know whether one could look forward in the near future to results at the genome-wide level for Neolithic individuals. At the time, the major international project with the goal of sequencing the DNA of the whole human genome (1990-2003) was producing year-after-year advances in methods and the equipment used in the laboratory. If one could bring together the required resources (this kind of lab work is very expensive) with a high level of virtuosity on the part of those leading the respective research teams (more on this below), then it was entirely possible that the genes of the first farmers might arrive on the scene sooner than most archaeologists expected in 2003. However, even as late as 2011, the important role that human genetics was about to play in the literature was still unclear as exemplified by the limited attention paid to this line of study in two major books: one on the Neolithic transition (Bocquet-Appel and Bar-Yosef 2008) and the other on the origins of agriculture (Price and Bar-Yosef 2011).

Indeed, the second decade of the 21st century would see the publication of a series of game-changing studies (e.g. Haak et al., 2010; Fu et al., 2012; Szécsényi-Nagy et al., 2014; Mathieson et al., 2015; Hofmanova et al., 2016; Olalde et al., 2019), which provide direct genetic support for the hypothesis of demic diffusion. More will be said about the respective studies in Section 4 of this chapter. Whenever genetic evidence is available both for the early Neolithic and for the Mesolithic in a given region of Europe, it shows that the genes are usually different. On the other hand, the genetic patterns for nearby regions of Europe in the Neolithic are commonly more similar to one another and even have links that can be traced back to Anatolia (Mathieson et al., 2015). To put it another way, direct genetic evidence does not provide support for the inference of continuity between the Mesolithic and the early Neolithic in most regions of Europe.

On the contrary, it implies that we have to move beyond the long running debate between demic diffusion and cultural diffusion in the literature of the 20th century (more on this in Section 3 of this chapter). Now direct genetic evidence clearly favors the hypothesis of demic diffusion. This does not mean that interactions between the first farmers and the last hunter-gatherers did not take place, however. Here it is worth recalling that this was our position from the start (Ammerman and Cavalli-Sforza 1973: 353; 1984: 116-19; see Section 2 below). Moreover, interaction continues to play its role in our recent work on "Modeling the role of voyaging in the coastal spread of the Early Neolithic in the West Mediterranean" (Isern et al., 2017; more will be said in Section 5 about how we incorporated them in modeling of the rapid rate of spread there; see Fig. 1 for a selection of early Neolithic sites in the Mediterranean world). Thus, what needs to be underscored, by way of introduction, is the shift in basic thinking that has taken place as the result of direct genetic evidence in the last ten years. In Section 2, the plan is to review some of the first steps that pointed us in this direction. Section 3 will return to the connection between indigenism, as a leading paradigm in the 20th century, and the claim that late foragers (Mesolithic people) were the protagonists of the spread of early farming in Europe. Section 4 will review briefly some of the recent genetic studies cited above. The last section will consider several questions that have not received the proper attention they deserve in the previous literature on the Neolithic transition in Europe.

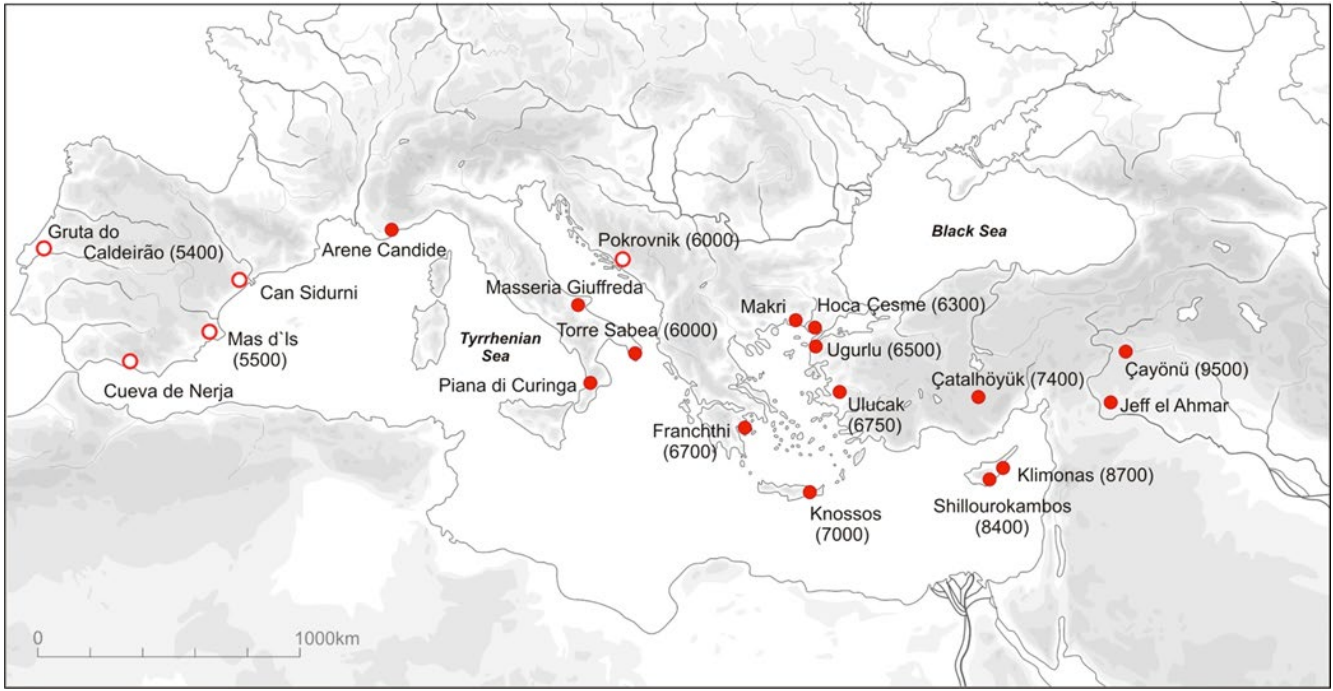

Figure 1. Map of the Mediterranean showing the location of a selection of early Neolithic sites. Obsidian is present at those with red dots.
[Ammerman 2014: fig. 3.]

## 2. First Steps

How did it happen that I became interested in the problem of the spread of early farming in Europe? And what made it possible for Luca and I to write the two papers that we did in 1971? Of course, there are no short or simple answers to questions of this kind. In the first chapter of *The Widening Harvest (*Ammerman 2003), I did say a few words about what we were trying to do when we met, but my main aim was to explain the purpose of the book. It is worth adding here a few more of the elements that led me in the direction of doing something quite new and different before returning to our first two papers and then commenting on the contribution of the Acconia Survey in our better understanding of the process of the relocation of Neolithic household and settlements on the landscape. Nothing would have ever happened if I had not chosen to go to London in 1967 and study environmental archaeology at the Institute of Archaeology. At the time, the Institute was an independent unit of the University of London with a focus on graduate studies.

By the way, I happened to have almost no formal training in archaeology when I arrived in London. I went to London to study ecology and pollen analysis with G. W. Dimbleby in the Department of Environment Studies; the Institute was then one of the few places in the realm of archaeology with such a department. And nothing would have happened without a fortuitous event that took place in London on May 18-19, 1968: a two-day conference on "The Domestication and Exploitation of Plants and Animals," which was held at the Institute of Archaeology and had the participation of specialists from around the world on the question of the origins of agriculture (Ucko and Dimbleby 1969).

I found this research topic to be one of great fascination, and the conference soon led me to change my field of study at the Institute. In short, there now arose the chance to do my doctoral dissertation under the supervision of John D. Evans, who had studied the prehistory of Malta for years and who had recently reopened the excavation of the Neolithic levels at Knossos on the island of Crete. He steered me in the direction of Italy and the Mesolithic period: that is, the late hunter-gatherers who lived at the time just before the first farmers. In Europe and America, there was in the 1960s an enthusiasm for the study of hunters and gatherers, and this interest now extended even to the less well-known Mesolithic period. Indeed, it had recently become fashionable in some circles for those interested in the origins of agriculture to attempt to trace independent pathways to plants and animal domestication in the context of late hunter-gatherers in several regions of Europe, including Greece, Romania and France. By the way, it is now well documented that pigs were first domesticated in the Near East and then moved westward as part of the Neolithic package (Laurent et al., 2019; subsequently, there was in Europe substantial gene flow from wild boar). Today there is also good evidence from the Franchthi cave in Greece for a quite rapid shift from the hunting of wild animals such as red deer to animal husbandry based on the herding of sheep (Munro and Stiner 2015). In contrast, the basic idea then was that the archaeologist should look for continuity between the Mesolithic and the early Neolithic in terms of faunal remains, lithic technologies and settlement patterns. To put it another way, the study of the spread of early farming from the Near East to Europe had fallen out of fashion by the late 1960s. For years, there had been no shortage of discourse that involved the use and abuse of migration and diffusion to explain it -- so why not have someone look into the case of Italy as well. Thus, I was sent out to the *Bel Paese* as part of the new urge to give Europe its own agricultural roots, which would ostensibly be taken back to the Mesolithic period.

There were two main lines of evidence that I set out to collect and study in Italy. The first one involved comparing sites of Mesolithic age with those dating to the early Neolithic in the different regions of the country based on their environmental settings on the landscape.  This would require visiting many sites and making observations on the ground.  In Italy, no comparative study of this kind had been done before, and the fieldwork represented a natural extension of my original reason for going to London to conduct environment studies in the context of archaeological sites.  But there was a major snag:  most of the known pre-Neolithic sites in Italy did not have a position on the map.   And this was entirely understandable, since many of them took the form of either a cave or else a rock shelter.  In short, there was no easy or reliable way for the Italian state to protect such sites from *clandestini*:  that is, from amateur archaeologists with a passion for digging on their own at them and thus damaging the sites. Accordingly, when professional archaeologists published reports on their excavations, they wisely chose not to give away the locations of their site.  The place where a site was said to occur in a report usually stood at a considerable distance from where the site was actually located.  In other words, it might take a fair amount of detective work to find a given site in the field -- before there was the chance to make observations on its environmental setting.  By visiting a large number of early sites and by making commitments to professional archaeologists in Italy not to reveal their sites' locations, much of this information has quietly remained buried in the pages of my thesis for years.  The second line of research concerns the analysis of the chipped stone tools that belonged to the lithic assemblages recovered at the respective sites.  Without going into the detail here, this led, in turn, to my work with Roy Hodson on the computer analysis of a comprehensive set of lithic assemblages in Italy, which was the topic of the paper that I gave at the Mamaia conference (Ammerman 1971).  More importantly, the meeting in the summer of 1970 gave me the chance to meet and talk with Luca for the first time.  In high school and as an undergraduate at the University of Michigan, I had acquired a sound background in mathematics.  Without it, I would not have been in contact with Hodson, and the meeting with Luca at a resort on the Black Sea in the summer of 1970 would never have happened.  All in all, it was a totally improbable run of events that led to our first contact and then collaboration.

In preparing for the seminar at the University of Pavia, I put together a list of the earliest Neolithic sites in the different countries of Europe that currently had radiocarbon dates and a map that showed their spatial distribution.  For me, it was time to return to the big picture that was emerging with respect to the spread of early farming in Europe and take another look at the question with those in the fields of genetics and biology who shared quantitative interests.  By November of 1970, I had come to realize that it would not be possible to trace an independent pathway to food production in Italy based on studying the Mesolithic there.  In terms of their lithic traditions, there was little evidence for continuity between the last foragers and the first farmers in their respective technologies and stone tool types.  And this likewise held when one compared the environmental settings of sites dating to the two periods.  The early Neolithic sites were commonly located in places with good potential for settlement on a year-round basis and for a subsistence strategy based on agro-pastoralism.  In contrast, what one had on the Mesolithic side were, for the most part, rougher, more limited and less favorable places, where campsites could be occupied only on a seasonal or part-time basis.  Moreover, there were almost no open-air settlements of Neolithic age in Italy where excavations had been taken down to the natural soil and yielded archaeological remains of Mesolithic age.  In other words, the working hypothesis that there was continuity between the last foragers and the first farmers in Italy was not panning out.  Looking back from where we stand today, my understanding of the situation in 1970, based on the rather limited evidence available at the time, turns out to be on the mark.  In a more comprehensive and sustained study of the question published thirty years later, Carlo Franco (2011) found little evidence for continuity between the late Mesolithic and the early Neolithic in Italy.  In retrospect, I happened to make my visit Pavia at just the right time, when I was drawing back from the notion of Mesolithic-Neolithic continuity.  It was time to look at the problem with fresh eyes.

My talk at the University of Pavia went well.  Luca and I then went on to exchange further ideas on new directions in this field of study, including the possibility of measuring the average rate of spread of early farming over Europe as a whole. Originally, the plan was for me to return to Rome on the day after the seminar.  But we kept on probing the question over the weekend at his house in Milan – to the dismay of Alba, his wife, who was apparently accustomed to taking it all in stride with unswerving patience.  Luca was now at the peak of his career and on his way to Stanford.  He had a quick and wide-range mind, and he seemed to recognize from the start that our collaboration would be productive.  By the way, he too had an interest in hunters and gatherers.  He had recently organized an expedition to the rain forests of central Africa to study the pygmies living there, and he was now making plans for a second trip (Cavalli-Sforza 1986).  In addition, he already had the intuition that major events in prehistory had left their "footprints" in the genes of present-day human populations.  Only a few highlights of our first steps can be considered in the space available here.  Our initial focus was twofold: on the overall pattern of the spread of early farming and on measuring its average rate of spread, as mentioned above.  In the case of Europe, this was not something that had been attempted in the previous literature.  Looking back, it happened to be just the right time to attempt such a measurement.  In 1946, Willard Libby at the University of Chicago proposed a new method for dating organic materials by measuring their content of carbon 14, a recently discovered radioactive isotope.  This initiated the so-called "radiocarbon revolution" in the dating of archaeological sites. By 1970, enough Neolithic sites had been carbon-dated, and they also had a wide spatial distribution on the map of Europe to make it possible to estimate the average rate of spread at the first level of approximation. The approach taken to the collection of the data (53 early sites with C-14 dates) and the treatment used in the regression analysis, which yielded a value of approximately 1 kilometer per year, were the subject of our first article. And even at that time, we were able to recognize that some regions of Europe such as the West Mediterranean and the Linear Bandkeramik culture (LBK) in central Europe had rates of spread that were faster than the average one for Europe as a whole (Ammerman and Cavalli-Sforza 1971: table 2). Within a few years, there were many more early Neolithic sites with radiocarbon dates, and this now meant that we could use 106 dated sites to produce a computer-generated map of Europe that shows the pattern of spread of early farming on the basis of a series of isochron lines (Ammerman and Cavalli-Sforza 1984: fig. 4.5). In 2005, the measurement of the rate was done again (with a total of 735 sites with C-14 dates), and it too yielded the same basic average rate as obtained in previous studies (Pinhasi et al., 2005).

Of no less interest in 1970 was how to explain what we were now trying to measure. As mentioned before and discussed at greater length in the first chapter of *The Widening Harvest* (Ammerman 2003: 6-11), diffusion was at the time out of favor as a way of explaining human behavior in the fields of human geography, anthropology and archaeology. Now if one even entertained the idea of using it, the assumption was that one was talking about cultural diffusion (Edmondson 1961). For decades, there had been the excessive use of terms such as "migration" and "colonization" whenever a scholar wrote about the movement of people. In short, what was not taken into account was that people have always been moving around on the landscape at the local level, as scholars working in the field of historical demography know quite well from the study of parish books that record births, marriages and deaths. In the case of small villages in the upper Parma valley during the 17th century, for instance, a person could be born at one village, then married in another one and finally buried at a third village. In short, human mobility or migratory activity can often be studied on landscapes in Italy over the course of the last five centuries and even more. Cavalli-Sforza was fully aware of this dimension of human population dynamics since he was the head of a group that pioneered genetic studies that drew upon consanguinity in the Parma valley (for references at that time, see the bibliography in Cavalli-Sforza et al. 2013). Thus, we realized in 1970 that what was called for was a better term than "migration" -- one that was broader in scope and that made due allowance for human mobility at the local and regional levels. During our first discussions, the new term that we now came up with was that of *demic* diffusion (Ammerman and Cavalli-Sforza 1973: 344). This was, in its own right, an important step in the Odyssey that we were embarking on.

The innovation that is much better known is, of course, the model of the wave of advance. During his studies at Cambridge University, Cavalli-Sforza was in regular contact with the geneticist R. A. Fisher, who put forward the model of "the wave of advance of advantageous genes" (Fisher 1937). In a nutshell, he brought together three elements in an equation that treats the growth and spread of a new gene in a population as the outcome of a diffusionary process. Without going into the details here, the first element concerns the initial growth rate of the new gene in the population; the second element involves the migratory activity associated with the population. Working together, they give rise to the radial rate at which the new gene is expected to spread or advance in the population (see Fig. 2).

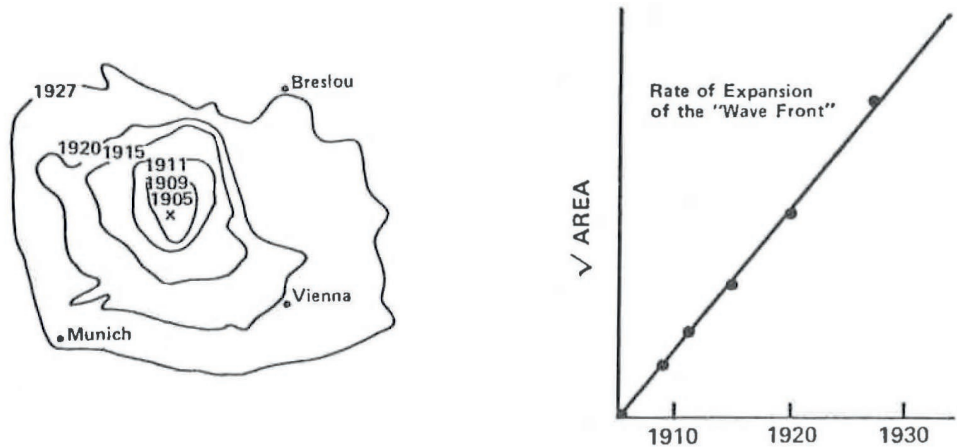

Figure. 2. Fisher's model of a population wave of advance is illustrated in this diagram. [Ammerman & Cavalli-Sforza 1984: fig. 5.3; after Skellam.]

This basic formulation was then taken up and further developed by Skellam (1951; 1973) in his work on the mathematical modeling of diffusionary processes in population biology: recall his classical example of the spread of the muskrat after its introduction in central Europe (e.g. Ammerman and Cavalli-Sforza 1984: fig. 5.3). This is a good place to say a few words about a common misconception with regard to the model in the literature on the Neolithic transition. It concerns how one thinks about the rate of spread when one uses the wave of advance model. No assumption is made that it advances at one fixed rate on the ground at all times and in all places (for more on this misconception, see Ammerman 2021: 5-6). This can be seen in the contour lines for the respective years of 1915 and 1927 on Skellam's map. They do not take the form of perfect circles. Instead, they have irregular shapes, "as might be expected in the case of an animal species spreading through a heterogeneous environment.  But if the law of averages is allowed to operate, regularities emerge" (Ammerman and Cavalli-Sforza 1984: 70).  The wave of advance is a model of population dynamics in the field of biology (where *constant* in a mathematical formula means *regular* on the ground) and not a model of chronology in archaeology.  In the case of the spread of early farming, we too are dealing with a real world:  one with a fair degree of local variation in time and space and not a spread that is simply moving forward with a uniform pattern on the ground.  In the case of early farming in Europe, the leading edge of the wave is, on average, advancing quite slowly at a distance of only about 25 kilometers per human generation.  An introduction to the model and how it can be used in studying the Neolithic transition is presented in our book (Ammerman and Cavalli-Sforza 1984; see also Ammerman and Cavalli-Sforza 1979).  Here it is worth mentioning the major contribution that Joaquim Fort then made to the formal development of the model as it relates to the specific context of archaeology (Fort and Mendez 1999).  Returning to Pavia in 1970, it is fair to say that we managed to get off to a good start in a matter of a few days.

Once I reached Stanford in 1972, there was much for those of us in the research group of Cavalli-Sforza to do.  In addition, I taught two courses in the new Program in Human Biology:  one on the transformation of human subsistence and the other on human evolution.  The Department of Genetics had its own library, and this gave me the chance to read almost everything that came out in the fields of human genetics, population biology and demography.  There was also the opportunity to carry out a simulation study of how population growth animates the rate of spread of first farmers in the context of settlement patterns such as those of the Linear Bandkeramik (LBK) in central Europe.  This gave me insight into how the three components of the wave of advance model could interact with one another on the ground.  During my second year at Stanford, there was the chance to visit the Aldenhoven Platte in Germany, where 17 LBK settlements were found over a 9 km long stretch of the stream called the Merzbach (Ammerman and Cavalli-Sforza 1984: fig. 3.5).  Over a 1.3 km stretch of the Merzbach, there was the complete excavation of the landscape in connection with large scale mining operation for brown coal, which produced a total of 160 long houses of LBK age (Kuper et al., 1974), providing good evidence for sustained population growth in the area (Ammerman and Cavalli-Sforza 1984: fig. 5.6; for more on our simulation study, see Ammerman 2022).

By this time, I also had a visiting position at the new Laboratory of Ecology at the University of Parma, which gave me the chance to keep up with the latest Italian literature on historical demography and more importantly to conduct archaeological fieldwork in Italy.  Then something quite unexpected happened in 1974.  Through the network of Luca's colleagues in Italy, Gianfranco Ghiara, the acting head of the new

University of Calabria, invited me to organize an archaeological survey with the aim of finding Neolithic sites in the toe of Italy. At the time, almost nothing was known about the Neolithic period in the region. Indeed, Bernabò Brea (1966: 43), a leading prehistorian, had recently noted that, while the Neolithic period was well known in the nearby regions of Apulia and Basilicata as well as on the island of Sicily, it had somehow skipped over Calabria. In effect, the survey was starting from a blank slate, and there was always the chance that we would return home empty-handed. On the other hand, the Calabria Survey gave me the chance to see what I could do on the more practical side in the field and not just in terms of modeling and theory. In making plans for the survey, Keith Kintigh, who was then doing graduate studies at Stanford in the new field of artificial intelligence, and I created the AMPRA survey game, which was then played both by a group of professors with an interest in statistics and by graduate students at Stanford taking a course in sampling design (Ammerman 1985: 3-4). This exercise proved to be highly productive and led us to develop an innovative strategy in the design of the Calabria Survey. On the basis of initial reconnaissance work, we first identified four areas of comparatively small size with different environmental settings and then began to cover each of them in a contiguous way so that comprehensive patterns of settlement were obtained. In the years between 1974 and 1976, all four areas produced a reasonable number of Neolithic sites, but the one called Acconia was clearly the richest (Ammerman 1985). There, within an area of less than 7 square kilometers, we were able to recover a total of 75 prehistoric sites by means of the intensive and repeated coverage of the landscape. And most of the sites were of Neolithic age. While we recovered some stone tools that went back to the time of the Middle Palaeolithic, there was nothing of Mesolithic age on the landscape at Acconia.

Thus, we soon learned that the Neolithic had not skipped over Calabria. All that one had to do was to look for the Neolithic sites in a more systematic way. Without going into the details of the Acconia Survey, we were able to find ten settlements of good size with impressed ware Neolithic pottery in the Stentinello tradition (dating from ca 7,900 to 7,300 years ago). At the settlement called Piana di Curinga (Fig. 3), we then went on to conduct a magnetometer survey, to make by hand a large number of cores and to carry out a series of excavations, which made it possible to document the presence of 48 buried wattle-and-daub houses of Stentinello age within an area of 2 hectares (Ammerman et al., 1988: fig. 2). Because of the 'Ndrangheta, the local name for the mafia in Calabria, it was not feasible to conduct at Acconia open-air excavations on a large scale of the kind made at Aldenhoven and Bylany in central Europe. Be that as it may, much the same level of local population density was now coming to light at Acconia in Italy as observed at Aldenhoven in Germany. In 1979, there was the good fortune at Piana di Curinga to excavate the house in Area H, which was well preserved by a fire and led to the recovery of a ton of baked fragments of daub (often with impressions of timbers and wattles in them) as well as whole ceramic vessels below the collapsed house.

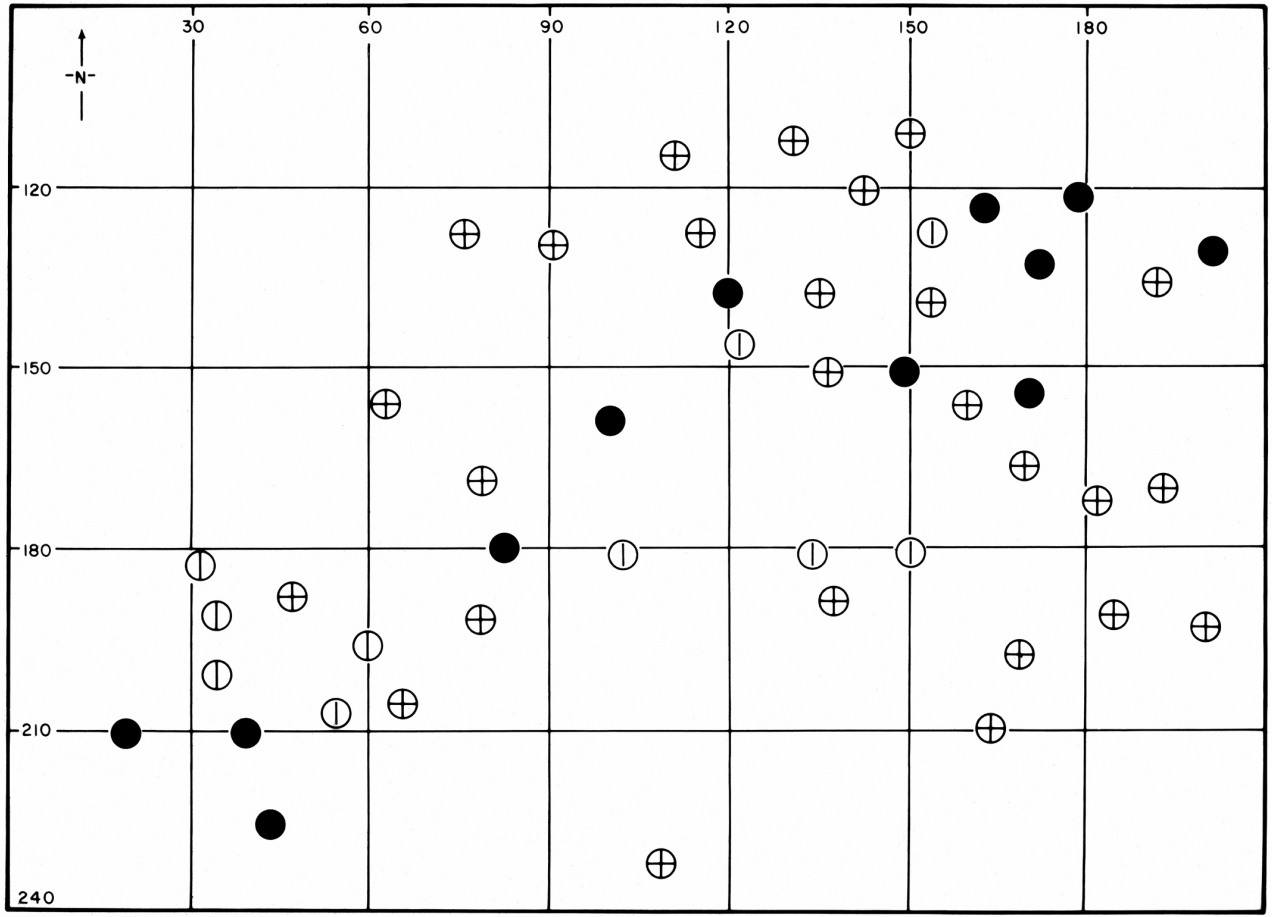

Figure 3.  The spatial distribution of the 48 wattle-and-daub houses of Stentinello age at the settlement of Piana di Curinga at Acconia (see Fig. 1 for its location). Circles with a cross indicate those cases (26) where a structure was first identified as an anomaly by the magnetometer survey and then confirmed by a hand boring. Solid circles represent those cases (12) where a house was tested by excavation following its identification by the magnetometer and borings.  Circles with a vertical line comprise those cases (10) where a structure was identified by means of hand borings alone (based on the recovery of fired daub).  The grid system is in metres.

What we learned at Aldenhoven and Acconia was that the relocation of one or more households at a given settlement to a new settlement was the key to understanding the human behavior that produced the Neolithic transition on the landscape.  Previously, the idea of relocation was rarely mentioned in Neolithic studies.  Now it was turning out to be at the heart of demic diffusion:  settlement relocation was "the induction to movement" (Skellam 1973) in the case of the first farmers.  From the perspective of human behavior, this is what they were choosing to do on the landscape:  to relocate from one settlement to the next one.  It was not by means of migrating from one region to the next that the spread of first farming actually happened.  Instead, it was by means of relocating from a settlement to a new one and by doing this over and over again that the Neolithic transition often took place on the ground.

## 3. The Paradigm of Indigenism

In order to gain a better understanding of why so many archaeologists in Europe, especially those in Britain, firmly believed to seeing the Neolithic transition as a matter of Mesolithic continuity, I came to realize that it was due to the paradigm of indigenism, which had become deeply ingrained in the social sciences in the second half of the 20th century.  The term *indigenism* is first mentioned in the preface to our book (Ammerman and Cavalli-Sforza 1984: xiv): "As part of the reaction against earlier diffusionism, a counter position – what for want of a better term might be called *indigenism* – has become popular in some quarters.  Sources of change and innovation are sought within indigenist populations or local cultural contexts."  In a short article in *Antiquity*, the concept of indigenism is briefly taken up again (Ammerman 1989: 165), as "an orientation in prehistory that would view societies and population as operating as closed, endogenous systems" and that "casts cultural change as a self-contained affair."  Several of the strands of thinking that gave rise to indigenism were identified as well. In the first chapter of *The Widening Harvest*, I would return to them at greater length (Ammerman 2003: 13-15; more on this below).  In fairness, it is worth noting that, in the 1980s, some of the archaeologists who were advocates for indigenism seemed to have reason on their side at the time.  There was, for instance, the idea that sheep had been locally domesticated in France based on fauna remains recovered from a few cave sites and rock-shelters in the Aude (Geddes 1987).  But when the evidence was reviewed using new and more rigorous approaches to taphonomy of the sites, this working hypothesis did not pan out.

In 1989, what I was trying to figure out was why an archaeologist like Marek Zvelebil had such a firm conviction that the Mesolithic was the protagonist of the Neolithic transition.  In fact, he believed so strongly in Mesolithic-Neolithic continuity -- that he had a hard time reading our book.  For example, he identifies our position as "the wave of advance hypothesis" (Zvelebil and Zevelebil 1988: 162).  But we have never used this expression. On the part of Zvelebil, there is confusion between two quite different things:  a model and a hypothesis.  In our publications, we have always drawn a clear distinction between the hypothesis of demic diffusion (where the form of the movement on the part of first farmers can take a number of different forms) and the wave of advance model. A model is a tool for thinking as well as a framework for addressing a problem (Ammerman 2003: 8): "The value of a model, as the economic historian David Landes (1969: 540) notes, is accordingly heuristic. It does not tell us what happened in the past; it helps us to discover and understand what may have happened." Here it is worth mentioning a second example of Zvelebil's belief in the paradigm. On one hand, he is able to acknowledge that there is good reason to think that demic diffusion did take place in a fair number of countries in Europe (among them in his view were Greece, Bulgaria, Romania, Hungary, Italy, Austria, Czechoslovakia, Poland, East and West Germany, Belgium and the Netherlands), and yet he considered them to represent "an exceptional situation" (Ammerman 1989: 164). In his view and that of many other British archaeologists, what they viewed as typical at the time was what they thought had happened in their own neck of the woods: that is, in northwestern Europe (the UK, Ireland and Scandinavia). Archaeologists working on the continent commonly saw things the other way around.

The belief in indigenism at that time is also reflected in *Problems in Neolithic Archaeology*, the book written by Alasdair Whittle (1988), in which the author avoids taking up the leading question in Neolithic studies at the time, the debate over the Neolithic transition, since this would mean that he would have to present the new alternative of demic diffusion. More recently, he has moved toward a more open and balanced position in the book that he co-edited with Bickle called *Early Farmers: The View from Archaeology and Science* (Whittle and Bickle 2014). Their opening chapter includes their study of LBK men and women who did relocate during their lifetimes quite often on the loess landscapes of central Europe, as seen on the basis of their bone chemistry (isotopes of strontium). In the book, there is also a chapter by Anna Széscényi-Nagy and co-authors that presents good evidence from Neolithic DNA for the LBK spreading by demic diffusion in Hungary (more on this in Section 4). On one hand, scientists in fields such as genetics, biology and the earth science were quite ready and willing to read the arguments put forward in our book when came out in 1984. On the other hand, many scholars in the fields of archaeology and anthropology who were true believers in indigenism encountered a mental roadblock when they attempted to read the book. In retrospect, there was, at the level of a paradigm shift, a tug-of-war between indigenism and science that was taking place in the minds of many archaeologists at the time. Given the results of recent genetic studies of human bones of Neolithic age, it is now clear that science was pointing the archaeologist in the right direction, while indigenism was holding back new knowledge as well as a better understanding of the Neolithic transition in Europe.

At this point, it is useful to turn to four strands of thinking that underlie the attraction of indigenism in our time.  When the term was first put forward in the preface of our book, no attempt was made to explore its intellectual roots.  I finally began to do this in 1996 when I was working on the history of the reconstructions of ancient Rome, as mentioned before, and when I was starting to make plans for the Wenner Gren Workshop in 1998.  In short, where does the idea come from?  And why does it have such a deep appeal today?  On the first strand that I recognized, let me quote from the introduction to *The Widening Harvest*: "To begin with, indigenism represents the working principle behind the majority of the studies in cultural and social anthropology since the second world war. In this field, a lone anthropologist would set out with a knapsack full of notebooks to chart the rules of organization of a remote, indigenous society.  Notably, indigenism is at the heart of the school of thought known as structuralism" (Ammerman 2003:13).  This term was coined by the anthropologist Claude Levi-Strauss to identify a method of applying models of linguistic structure to the study of a society and, in particular, to its customs and myths.  In structuralism, emphasis is placed on cultural difference and the importance of respecting it.  Change over time is not a leading concern, and if it does occur, it is held to have its source within a given society.  In other words, both structuralism and indigenism speak of cultures as if they are autonomous.

The second strand, which is more directly related to prehistoric archaeology, is linked with nationalism.  In the 1960s and 1970s, there was the emergence of the idea that each nation in Europe was entitled to have agricultural origins of its own.  As mentioned before, a series of chauvinist claims were made for the local domestication of a plant or an animal in Greece, Romania and France, but none of them has withstood the test of time.  In Europe, archaeology is organized essentially along national lines.  This carries with it the unequivocal aspiration that each nation's archaeology should advance the identity of its country.  Thus, each nation's archaeology has the agenda of tracing its own distinct and preferably independent course of history over time (Ammerman 2003: 14).

A third strand contributing to indigenism is the sense of apprehension that we all feel when it comes to the displacement of people in the post-Holocaust world.  Our views on the movement of people have been redefined by the events of the Holocaust.  In the 20[th] century, there had already been enough of this sort of thing in the 1940s, and then there came along in 1999 more forced displacements of people in the case of Kosovo and East Timor as well.  And it has not stopped in the present century.  Of course, this runs counter to the sensibility of the post-modern world.  The concept of indigenous people, however, contains its own paradox. Today many social groups that wish to identify themselves in this way have actually experienced one or more relocations in recent historical times, and they are no longer autochthonous.

The fourth strand involves what amounts to a post-colonial involution. Here the basic aim is to rewrite the past: that of societies in prehistory as well as one's own history in more recent times. The appeal of indigenism seems to be more pronounced in nations with a long colonial experience. "In adjusting to a world that has changed, there is the need to realign one's identity. We live in an era of empires giving way to the conscious rediscovery of ethnic identity and to the natural right of self-determination. In this context, there is a strong urge to affirm that one is in control of one's own native past. By way of involution, when it comes time to look outward again and consider other societies, the archaeologists tend to project indigenism on the rest of the world as well" (Ammerman 2003: 14). Together these four strands produce a complex of values and attitudes with a firm hold on our time. Thus, there is more at stake, when it comes to the Neolithic transition, than just what we know about how first farming began in Europe. If our formulation of the question is correct, as now appears to be the case in light of direct genetic evidence from human bones of Neolithic age, then the Neolithic transition represents a major challenge to the paradigm of indigenism, which has been in vogue for the last 50 years.

## 4. The Direct Genetic Evidence and its Implications

The purpose of this section is to present, in telegraphic form, a selection of publications that illustrates the rapid advances that have taken place since 2005 in the direct genetic analysis of the bones of early Neolithic individuals in various part of Europe and the new support that they now provide for the hypothesis of demic diffusion first put forward in the 1970s. There is not space here to develop a more systematic review. Most of the reports are short ones that have come out in high-profiled journals where their publication is truly merited. On the other hand, this means that discourse is often synoptic and at times even modest when it comes to explaining the processes that were involved in the Neolithic transition (more on this below). In selecting the studies selected here, part of the aim was to cover broadly the various regions of Europe. The seven reports are given in chronological order: each in a separate paragraph with a short heading at the start, next its title and then a few brief comments.

*A false Start.* We begin with the ill-fated report by Haak and co-authors (2005) called "Ancient DNA from the first European Neolithic farmers in 7500 year-old Neolithic sites." In terms of methodology and its basic results, it was clearly a breakthrough article at the time, and the editors at *Science* were eager to publish it. However, it was not carefully reviewed on the archaeological side (Ammerman et al., 2006). Without going into the details here, the authors did not have a good grasp of the chronology of their samples in central Europe. In fact, they seem to be unaware that one of their substantive conclusions is contradicted by their own data. The editors of Science were not happy to hear from us (mere archaeologists), but they dutifully published our comment online in a place where few readers at that time would find it.

*The real Thing*. The publication in 2005 was a setback for Haak and his team of geneticists, but they soon refined and extended their research in an article entitled "Ancient DNA from European early Neolithic farmers reveals their Near Eastern affinities," which was published in *PLoS Biology* (Haak et al., 2010). Drawing upon seminal work that came out the previous year (Bramanti et al., 2009), this study now provided good support for demic diffusion and ushered in a decade of further direct genetic studies that would confirm their result. On the archaeological side, discourse on the processes that produced the pattern of affinities was brief and uneven in character.

*Making the most of what is available*. The next report is an article in *PLoS One* by Fu and co-workers (2012) called "Complete mitochondrial genomes reveal Neolithic expansion into Europe." It too was music to my ears and those of Cavalli-Sforza. One of the co-authors of this paper was Svante Pääbo, the researcher at the Max Plank Institute in Germany who led the celebrated study of DNA in Neanderthals. As a virtuoso at studies of this kind, he must have been proud of how much the team could learn about ancient population history from the limited ancient genetic data that were available in this case study. One of the conclusions reached is that pre-Neolithic hunter-gatherers in Europe contributed only about 20% of the genes to the Neolithic expansion.

*On the Hungarian Plain*. This brings us to the chapter by Szécsényi-Nagy and co-authors (2014) on "Ancient DNA evidence for a homogeneous maternal gene pool in sixth millennium cal BC Hungary and the Central European LBK," which appeared in the book edited by Whittle and Bickle (2014). Their study now confirms in greater detail for Hungary what Haak et al. (2010) had found a few years before. Their chapter is, of course, much longer than the other articles included in this section, and it exhibits a higher level of discourse on the Neolithic transition. As mentioned before, the authors' working hypothesis, at the start of the study, is that early Neolithic genes would show east to west patterns of regional variation in this part of Europe. What they find instead is that direct genetic evidence is essentially homogeneous in space. In order to achieve such homogeneity, there has to be the active relocation of people not only forward to the west but also back towards the east, as we learned from our simulation study of LBK settlement systems year ago (Ammerman and Cavalli-Sforza 1979: 288-91; 1984: 113-16).

*A major Breakthrough*.  The next report, which was published in *Nature*, was written by Mathieson and co-authors (2015), and its title is "Genome-wide patterns of selection in 230 ancient Eurasians."  The article breaks new ground in several ways: (1) there are samples from the western part of Anatolia for the first time; (2) there is genome-wide evidence for a larger sample size than seen in the literature before; and (3) there is an attempt to search for patterns of natural selection in ancient DNA. There is an even better documentation from direct genetic evidence that the spread of early farming began in the Near East and then continued to move over Europe with little evidence for contributions from the Mesolithic side in most regions.  Here the virtuoso behind the scenes is David Reich at the Medical School of Harvard University, whose talents in the laboratory make it possible to run so many samples.  Again, the weak suit of the article has to do with the archaeological interpretation of the genetic data and discourse on natural selection.  Language has become too compressed to the point where it no longer expresses what is really at stake.

*The Aegean Connection*. The next stop is the report by Hofmanova and co-workers (2016) called "Early farmers from across Europe directly descended from Neolithic Aegeans," which came out in the *Proceedings of the National Academy of Sciences.* Based on a small number of samples (7), this study shows that first farmers from the east did pass through the Aegean region on their way to the west.  With the usual Greek spin on things, the article has its chauvinistic touch:  not much is said about western Anatolia or Turkey -- the previous step in the spread of early farming (Mathieson et al., 2015).  Some allowance is made for a Mesolithic contribution to the Neolithic transition in the Aegean but it involves rather slim evidence (only 2 samples from the site of Theopetra).

*The Iberian Peninsula*.  We move to the west end of the Mediterranean in the recent report by Olalde and co-authors (2019) on "The genomic history of the Iberian Peninsula over the past 8000 years" (see also Valdiosera et al., 2018; for neighboring France, see Brunel et al., 2020; Rivollat et al., 2020).  This article came out in *Science*, and it represents another large-scale study carried out under the guidance of Reich, the *impresario*.  As one would expect on the basis of previous studies in Iberia on a smaller scale, this part of Europe now joins the club of the first farmers who spread over Europe.  What is new in Spain and Portugal is placing this development in the wider context of the genomes of human populations that subsequently lived there.  In the brief discussion of the early Neolithic, the role of demic diffusion is taken almost for granted and with a hint of boredom:  that is, there is little that is new to say on the topic. Be that as it may, the article gives a synopsis – not really a history – of the changing patterns of genes linked with the various cultural periods running up to the present time.

From a diachronic perspective, the study of human genes in Europe and the big event of the Neolithic transition turned out to be highly productive during the second decade of the present century. And there is still more to come in the next ten years from the direct study of ancient DNA. At this point in time, it is premature to say where it will all lead, but the weight of the new genetic evidence supports the hypothesis of demic diffusion as the best way to explain the Neolithic transition in Europe. As mentioned before, the analysis of ancient DNA is very demanding and expensive. Thus, it takes high-profile publications in journals such as *Science*, *Nature* and *PNAS* for a laboratory to maintain its high level of funding and to keep major grants rolling in. However, publication in a major journal is not without problems of it own. The article has to be short (4 to 6 pages in all), and one-half of the space or more goes to the abstract, the tables, the figures, the methods and the references. In short, there is not much room for the text itself, which means that archaeological discourse commonly suffers. In fact, there is often backsliding in the use of language in the compressed texts rewritten by editors. The reader keeps running across old terms such as "migration" and people "migrating" over Europe: that is, just what we attempting to move beyond in 1971. I have talked with geneticists such a David Reich about this, but they tend simply to nod their heads. What this means is that archaeologists will have to insist on higher levels of archaeological discourse. This may call for separate, parallel articles or chapters with more emphasis on archaeological: to avoid the false start mentioned above and to move in the direction of the case study on the Hungarian Plain (Szécsényi-Nagy et al., 2014).

## 5. Discussion

At Stanford in the early 1970s, Samuel Karlin, whose math skills made a major contribution to early attempts at genome sequencing, once opened a seminar by saying that scholars often do not work on a problem long enough. His sage advice has clear resonance when it comes to our work on the Neolithic transition over the years. In this final section, the plan is to discuss a few new directions in thinking about the problem and considering some of the new work that is called for. To begin with, it is fair to say that we have come a long way since the first study on the spread of early farming in Europe fifty years ago. However, as seen in the previous section, there is still the challenge of striking a good balance in the dialogue between the archaeologist and the geneticist in the context of a rapidly developing field of interdisciplinary study. This was easier to do in the 1970s, when advances were taking place at a slower pace, than it is today.

From the time of our first attempt to measure the rate of spread, there was an awareness that early farming moved forward somewhat faster in two parts of Europe: the Western Mediterranean and the LBK (Ammerman and Cavalli-Sforza 1971: table 2). Both of them are, by the way, located on the west side of the map of Europe. Today it is well known that the rate of spread was slow in the eastern Mediterranean (Ammerman 2011b) and much faster in the western Mediterranean (Isern et al., 2017). By bringing this realization together with recent gains in knowledge on the earliest voyages to the island of Cyprus (see various chapters in Ammerman and Davis 2013-2014), a new research topic of considerable interest has emerged. At this point, it is useful to say a few more words about the question of early voyaging to Cyprus, the large offshore island at the east end of the Mediterranean. Even as late as 2003, there was not much reliable evidence for pre-Neolithic voyages to the various offshore islands in the Mediterranean Sea. Indeed, conventional wisdom held for years that hunters and gatherers in the Mediterranean were reluctant seafarers and their sites dating to the time before the Neolithic period were hard to find on the Mediterranean islands (Cherry 1990). Thus, in the 20[th] century, a scholar who set out to study the question of early seafaring in the Mediterranean world commonly took the Neolithic period as the point of departure.

In 2003, I decided to go out to Cyprus and look for the missing pre-Neolithic sites on the island. By taking a new approach to reconnaissance work there, we soon found sites going back to the years before the Neolithic at Aspros, Nissi Beach and eight other places on coastal formations of aeolianite around the island (Ammerman 2010: 86-88). This meant that late foragers in the Levant were already making crossings to Cyprus during the cold snap known as the Younger Dryas, a time of heightened mobility at the end of the Pleistocene (Ammerman 2010: 82-83). While my project on Cyprus did not directly concern the Neolithic transition, it gave me a window on what was then coming to light in the work t the PPNB settlement of Shilourokambos (Guilaine et al., 2011) as well as the PPNA settlement of Klimonas (Vigne et al. 2017). At Shilourokambos, there is good evidence that the PPNB (Pre-Pottery Neolithic B) had reached the island from the mainland around 10,400 years ago -- with more or less the full Neolithic package, including domesticated sheep. For our present purposes, the sheep are of particular interest since they, like other ruminants, are prone to life-threatening diseases when they rest in a recumbent position for a sustained period of time. This would have been the case for a small boat with several sheep in it, which attempted to make a slow, paddle-driven crossing from the mainland to Cyprus during the PPNB (Vigne et al., 2013). Accordingly, Jean-Denis

Vigne and his colleagues in France had made the argument that PPNB crossings to Cyprus, in order to be successful, must have occurred at a somewhat faster pace: that is, use was made of small boats with rudimentary sails.

If this interpretation is correct, it has far-reaching implications for how we view the Neolithic transition. In a nutshell, it suggests that maritime technology had already reached the stage of development some 10,400 years ago when one could envision, in theory, the Neolithic package moving from Cyprus to the heal of Italy quite rapidly: for instance, in a span of 500 years or some 18 human generations. Reality was on a totally different page, however. Calibrated radiocarbon dates indicate that early farming eventually made its way to southern Italy only by 8,200 years ago. In a world where voyaging had deep time roots and where maritime technology had apparently evolved beyond its paddle-powered infancy, why did it take more than 2,000 years for early farming to move from Cyprus to Italy? This is what I have called "The paradox of early voyaging in the Mediterranean and the slowness of the Neolithic transition between Cyprus and Italy" (Ammerman 2011b).

To cut to the chase, it is of no less interest to extend this counter-factual line of argument to the Mediterranean as a whole. As can be seen in Figure 1, there is evidence that the spread of early farming arrived in Portugal around 7,400 years ago. In other words, it spread all the way from the heal of Italy to Portugal in only about 800 years. However, even when due allowance is made for a much faster rate of spread in the west Mediterranean (Isern et al., 2017), the question to ask is why did it take some 3,000 years for first farming to travel from Cyprus to Portugal in the context of early voyaging and voyagers who were actively engaged in the exchange of obsidian from their sources on Greek and Italian islands (Ammerman 2010: 83-86). To put it another way, the Neolithic transition somehow managed to take some 110 human generations to travel from east to west across the Mediterranean world, where boats and seagoing formed an integral part of the way of life at the time. When we step back and look at the big picture in this way, we realize that it could have happened much faster. This is then one of the new directions for the next generation to explore.

This chapter is not the place to discuss at length the paradox in the eastern Mediterranean and the wide range of factors that may help to explain why it took so long for the first farmers to reach Portugal. Only a few suggestions will be put forward briefly here. The different lines of explanation range from outbreaks of disease and their negative impact on the growth rate of human populations to the length of the time-delay in the relocation of settlements and from climate change to the perils of the sea and the loss of small boats during storms. In the latter case, one is dealing with a rare event that was more or less random in character.  With regard to a climate event such as the abrupt collapse of the Laurentine ice sheet at ca. 8,200 years ago (e.g., Berger and Guilaine 2009; Ammerman 2011b: 39), we are still at an early stage in learning about its nature, its duration and its local effects in different regions of Europe. Here emphasis will be placed on the demographic side of the story, since it entails processes that were always in progress at any one time or place during the spread of early farming.

From where we stand today, early Neolithic technology in the Mediterranean world – with its boats and people who knew how to use them – seems to offer considerable potential. But demography was not always in a position to take good advantage of it.  This was the situation in the Levant, Turkey and the Balkans where various cultures had the tradition of living on tells or mounds.  This meant that households lived in close proximity to one another on trash and building materials that had accumulated over time.  If there happened to be the outbreak of a common communicable disease at a mound, there was a good chance that it would progress through the closely spaced local community.   In short, mounds tended to foster episodes of high mortality, which worked against the growth rate of the local population. Again, from the viewpoint of epidemiology, those who occupied a mound were in a poor position to face a new deadly virus – a rare event that might happen every few generations with a devastating impact on the local population.  To make matters worse for the rate of spread of early farming, the time-delay involved in relocating from a given mound to a new on was a comparatively long one.  Since its length could last for several generations or even more in the case of mounds, it too acted to slow down the diffusionary process on the ground (as mentioned in Section 2).  While living on a mound had, of course, its positive social and cultural aspects, this form of settlement was not conducive to the rapid spread of first farmers.

In comparison, the situation was different in the western Mediterranean where mound sites of Neolithic age are not found in the archaeological record. Instead, the first farmers there have settlement patterns that take a more open and dispersed form.  In central Europe, a parallel shift away from mounds and towards a more broad-based and loose-knit way of living on the landscape is seen in the case of the LBK.  In terms of demography, one of the advantages of a dispersed pattern of settlement is a better chance of coping with vectors of disease:  both those of the everyday sort and in the rare event of a novel virus.  Of course, both of them will happen in the west from time to time.  But their effects, on the whole, will be more attenuated there.  The other positive feature of a dispersed settlement pattern is the shorter length of the time-delay in the process of relocating from a former place of habitation to a new one.  In short, it should come as no surprise then that the average rate of spread for the west Mediterranean as well as the LBK is several times faster than the mound-based one in the eastern Mediterranean. The marked contrast in the rate between the east and the west begins to make more sense. Only a rapid sketch of the differences between the east and the west in the Mediterranean world is given here. In effect, mound-based habitation applied its cultural brakes to demography in the east. Then dispersed patterns of settlement offered demography a new lease on life in the west. The sustained population growth that now ensued there, in combination with voyaging along the coast, produced the much faster spread of the Neolithic transition in the west (Isern et al., 2017). Once again, it will be the task of scholars in the next two decades to probe the depths of this new direction.

Another topic of interest that has not received enough attention in the literature concerns the paradox that is emerging on the Mesolithic side in recent studies of ancient DNA. In the direct study of genes from human bones, there are very few Mesolithic individuals recognized so far in those places where early farming initially spread: that is, those places on the landscape that were the most favorable ones in environmental terms for early forms of agro-pastoralism. In addition, Mesolithic genes are seldom found among those who lived and were buried in the context of the earliest Neolithic sites in a given area. Thus, the state of the evidence is rather disappointing for those archaeologists who had long hoped to find more in the way of interaction between final hunter-gatherers and initial farmers at the time of the Neolithic transition. On the other hand, there are sites that date to the middle Neolithic and even to the later Neolithic, which are located in places with less potential for early farming, where a few individuals are occasionally identified as having Mesolithic genes. In short, while Mesolithic genes appear to have gone missing at the start of the Neolithic transition, there is then subsequent evidence for their persistence in more recent Neolithic contexts, even if no attempt is made to explain this puzzle in the short articles that appear in leading science journals (e.g., Olalde et al., 2019). To put it another way, Mesolithic genes did manage to persist, even though they were rarely identified in those places where early farming first spread. Here it is worth adding that after the initial spread of first farming to the most attractive places on the landscape (in the eyes of the first farmers), there was then the on-going expansion of the Neolithic that filled in less favorable parts of a given region over time (for more on this question, see Ammerman 2022).

One way to account for this puzzle is that late foragers chose to keep their distance from those places where the farmers initially settled on the landscape. They now preferred to live elsewhere in the region. Accordingly, the foragers spent less time in forested areas with fertile soils that were of primary interest to first farmers, who were prepared to burn down trees to open up the landscape. It is worth recalling that hunters and gatherers simply walk away from a problem as a common response to conflict resolution. This adaption fits in well with their mobile way of life. The suggestion then is that groups of late hunter-gatherers in the Mediterranean preferred to stay away from first farmers and led a life of co-existence in other spaces in the same region. How long did this co-existence last? It probably varied from one region to the next, and this is a question that calls out for further study. In the case of Pygmies and Bantu farmers living in central Africa in our time, we know that it could persist for centuries (Cavalli-Sforza 1986). In the western Mediterranean, there is evidence for the side-by-side co-existence of the late Mesolithic and the early Neolithic in the estuaries of the Tagus and the Sado in Portugal that lasted at least for a number of generations (Zilhão 2001; 2011).

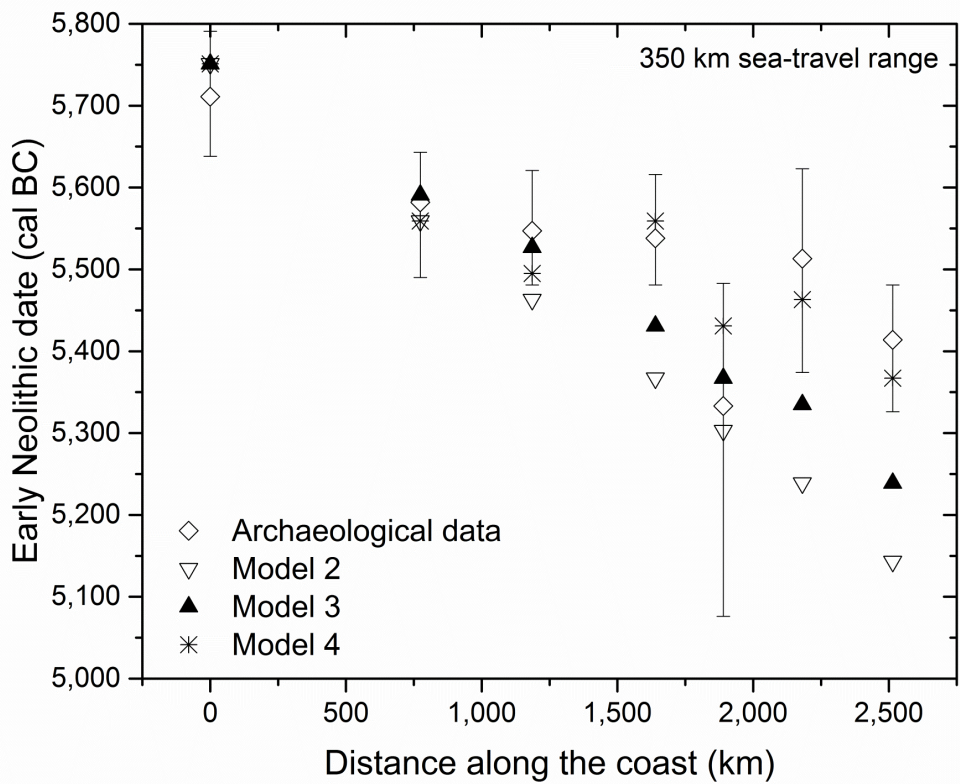

Figure 4. Early Neolithic dates plotted against distance along the coast from Arene Candide in Liguria (for its location see Fig. 1), which is displayed for the archaeological evidence and three simulation models. Model 4, which takes a leapfrog approach to site relocation, yields the best fit with lower voyaging ranges (from Isern et al. 2017: fig. 3).

Archaeologists who study the Mesolithic period are inclined to take the view that late foragers must have been involved in the story of the Neolithic transition. As mentioned before, there were even those in the last century who wished to see the foragers of the Mesolithic as the protagonists of the Neolithic transition. In the due course of time, most students of the Mesolithic became more cautious about this idea. And there were others, including Ammerman and Cavalli-Sforza (1984), who liked the idea of mutualism between late foragers and first farmers. Indeed, this is what we proposed in a recent article called "Modeling the role of voyaging in the coastal spread of the early Neolithic in the west Mediterranean" (Isern et al., 2017). For example, in light of relocations over considerable distances (according to the best fit model seen in Fig. 4), we think that local hunter-gatherers may well have acted as guides in new places where first farmers lacked local knowledge and that they also took part in voyages undertaken by first farmers. For a young adult male on the Mesolithic side, the real fascination of what was happening around them had less to do with the production and consumption of Neolithic foods and more to do with the technology of boats and the adventure of voyaging on the sea. This makes, of course, a good story for the archaeologist to tell, and it may well turn out to be the correct interpretation.

What the archaeologist is not supposed to do is to consign the lives of late foragers to the uneventful scenario of keeping their distance from the first farmers. Such a pro-active orientation is embedded in the concept of *Neolithization* itself. In effect, those leading a Mesolithic way of life are encouraged to play a positive role in one way or another in the Neolithic transition. What foragers are not supposed to do is to turn their backs on early farming and walk away from it. In retrospect, few archaeologists were reready and willing to let hunter-gatherers be just hunter-gatherers. An exception here is João Zilhaão (2011), who puts forward a critique of the concept of Neolithization for this reason and who makes allowance for "active rejection"-- as one of the ways in which a Mesolithic society could respond to the newcomers and their Neolithic package. Of course, there has been much debate on this issue in the literature. However, up until quite recently, there was no real way of testing the hypothesis of "stealthy rejection" – perhaps a better name for this concept. There are recent advances in genealogical reconstruction based on the analysis of ancient DNA (Racimo et al., 2020), which may make it possible to answer this question. Obviously, this kind of genetic work is quite demanding, and it may take some time before a number of studies are done properly. Indeed, there is a French study that recently came out (Rivollat et al., 2020) that makes the claim for admixture between first farmers and local hunter-gatherers taking place in the seventh generation after the initial arrival of farming in southern France. This would mean a time delay of around 200 years before Meolithic genes make their first appearance in the context of the earliest Neolithic sites there (for a more cautious reading of the genetic evidence in Southern France, see Brunel et al., 2020). In short, the prospect of sorting out this contentious matter is, in all likelihood, no longer beyond our reach. However, it will take time and concerted effort to carry out this demanding line of investigation based on a number of samples of high quality. In addition, further work needs to be done on the question of the relative importance of demic diffusion and cultural diffusion in the overall spread of early farming in Europe. On the basis of genetic evidence, Fu and co-authors (2012) attribute at most 20% to the Mesolithic side. On the basis of mathematical modeling, Fort (2012) comes up for a value of about 60% for demic diffusion. Today he regards this value as far too low. In the coming years, there will be the chance to refine such estimates for Europe as a whole and to take studies of this kind even down to the regional level as well. We have come a long way since 1970, and there is, of course, still much to do.

## References

Ammerman, A.J. 1971. A computer analysis of Epipalaeolithic assemblages in Italy. In: R.H. Hodson, D.G. Kendall, D.G. and P. Tautu (eds.). *Mathematics in the Archaeological and Historical Sciences*, 133-37. Edinburgh.

Ammerman, A.J. 1985. *The Acconia Survey: Neolithic Settlement and the Obsidian Trade.* Institute of Archaeology, University of London, Occasional Publication 10. London.

Ammerman, A.J. 1989. On the Neolithic transition in Europe: a comment on Zvelebil & Zvelebil (1988). *Antiquity* 63: 161-65.

Ammerman, A.J. 2003. Introduction. In: A.J. Ammerman and P. Biagi (eds.), *The Widening Harvest. The Neolithic Transition in Europe: Looking back, looking forward.* Archaeological Institute of America, Colloquia and Conference Papers 4, 3-23. Boston.

Ammerman, A.J. 2010. The first Argonauts: Towards the study of the earliest seafaring in the Mediterranean. In: A. Anderson, J.H. Barrett and K.V. Boyle (eds.), *The Global Origins and Development of Seafaring*, McDonald Institute Monographs, 81-92. Cambridge.

Ammerman, A.J. 2011a. Relocating the center: a comparative study. In: N. Terrenato and D.C. Haggis (eds.), *Questioning the Neoevolutionist Paradigm*, Oxford Books, 256-72. Oxford.

Ammerman, A.J. 2011b. The paradox of early voyaging in the Mediterranean and the slowness of the Neolithic transition between Cyprus and Italy. In: G. Vavouranakis (ed.), *The Seascape in Aegean Prehistory*, Monograph of the Danish Institute at Athens 14: 31-39. Athens.

Ammerman, A.J. 2021. Returning to the rate of spread of early farming in Europe: comment on the article by Manen et al. (2019) in Radiocarbon. *Radiocarbon* 63:1-9.

Ammerman, A.J. 2022 (in press). The transition to early farming in Europe. In: S. Bergin and S. Pardo-Gordo (eds.), *Simulating the Transition to Agriculture*. Springerh

Ammerman, A.J. and P. Biagi (eds.). 2003. *The Widening Harvest. The Neolithic Transition in Europe: Looking back, looking forward.* Archaeological Institute of America, Colloquia and Conference Papers 4. Boston.

Ammerman, A.J. and L.L. Cavalli-Sforza. 1971. Measuring the rate of spread of early farming in Europe. *Man* 6: 675-88.

Ammerman, A.J. and L.L. Cavalli-Sforza. 1973. A population model for the diffusion of early farming in Europe. In: C. Renfrew (eds.), *The Explanation of Culture Change*, 343-57. London.

Ammerman, A.J. and L.L. Cavalli-Sforza. 1979. The wave of advance model for the diffusion of early farming in Europe. In: C. Renfrew and K.L. Cooke (eds.), *Transformation: Mathematical Approaches to Culture Change*, 275-93. New York.

Ammerman, A.J. and L.L. Cavalli-Sforza. 1984. *The Neolithic Transition and the Genetics of Populations in Europe*. Princeton.

Ammerman, A.J. and T. Davis (eds.). 2013-2014. *Island Archaeology and the Origins of Seafaring in the Eastern Mediterranean.* Proceedings of the Wenner Gren Workshop held at Reggio Calabria on October 19-21, 2012. *Eurasian Prehistory* 10-11.

Ammerman, A.J., Pinhasi, R. and E. Banffy. 2006. Comment on ancient DNA from the first European farmers in 7500 year-old Neolithic sites. *Science* 312 (5782): 1875, doi: 10.11261/science.1123936.

Ammerman, A.J., Shaffer, G.D. and N. Hartman. 1988. A Neolithic household at Piana di Curinga, Italy. *Journal of Field Archaeology* 15: 121-40.

Berger, J.F. and J. Guilaine. 2009. The 8200 cal BP abrupt environmental change and the Neolithic transition: a Mediterranean perspective. Quaternary International 200 (1-2): 31-49.

Bernabò Brea, L. 1966. *Sicily before the Greeks*. London.

Bocquet-Appel, J.P. and O. Bar-Yosef (eds.). 2008. *The Neolithic Demographic Transition and its Consequences.* Springer, Netherlands.

Bramanti, B. et al. 2009. Genetic discontinuity between local hunter-gatherers and central Europe's first farmers. *Science* 326: 137-40.

Brunel, S. et al. 2020. Ancient genomes from present-day France unveil 7,000 years of its demographic history. *Proceedings of the National Academy of Sciences USA*, doi/10.1073/pmas.1918034117.

Cavalli-Sforza, L.L. 1986. *African Pygmies.* Orlando.

Cavalli-Sforza, L.L., Menozzi, P. and A. Piazza. 1994. *The History and Geography of Human Genes.* Princeton.

Cavalli-Sforza, L.L., Moroni, A. and G. Zei. 2013. *Consanguinity, Inbreeding and Genetic Drift in Italy (MPB-39).* Princeton.

Cherry, J. 1990. The first Colonization of the Mediterranean islands: a Review of recent Research. *Journal of Mediterranean Archaeology* 3: 145-221.

Edmondson, M.S. 1961. Neolithic diffusion rates. *Current Anthropology* 2: 71-102.

Fisher, R.A. 1937. The wave of advance of advantageous genes. *Annals of Eugenics, London* 7: 355-69.

Fort, J. 2012. Synthesis between demic and cultural diffusion in the Neolithic transition in Europe. *Proceedings of the National Academy of Sciences USA* 109 (46): 18669-73.

Fort, J. and V. Mendez. 1999. Time-delayed theory of the Neolithic transition in Europe. *Physical Review Letters* 82: 867-70.

Franco, C. 2011. *La fine del Mesolitico in Italia.* Società per la Preistoria and Protostoria della Regione Friuli-Venezia Giulia, no. 13. Trieste.

Fu, Q., Rudan, P., Pääbo, S. and J. Krause. 2012. Complete mitochondrial genomes reveal Neolithic expansion into Europe. *PLOS One* 7 (3): 332473.

Geddes, D. 1987. Subsistence et habitat du Mésolithique final au Neolithique moyen dean le basin de l'Aude (France). In: J. Guilaine et al. (eds.), *Premières Commuanautés paysannes en Méditerraneé occidentale*, Éditions du CNRS, 200-207, Paris.

Guilaine, J., Briois, F. and J.D. Vigne (eds.) 2011. *Shillourokambos: un établisssement néolithique pré-céramique à Chypre: les fouilles du secteur 1.* Errance-École Française d'Athénes. Paris.

Haak, W. et al. 2005. Ancient DNA from the first European farmers in 7500 year old Neolithic sites. *Science* 310: 1016-18.

Haak, W. et al. 2010. Ancient DNA from European early Neolithic farmers reveals their Near Eastern affinities. *PLoS Biology* 8, e10000536.

Hodson , R.H., Kendall, D.G. and P. Tautu, (eds.). 1971. *Mathematics in the Archaeological and Historical Sciences.* Edinburgh.

Hofmanova, Z. et al. 2016. Early farmers from across Europe directly descended from Neolithic Aegeans. *Proceedings of the National Academy of Sciences USA* 113 (25): 6886-91.

Isern, N., Zilhão, J., Fort, J. and A.J. Ammerman. 2017. Modeling the role of voyaging in the coastal spread of the Early Neolithic in the West Mediterranean. *Proceedings of the National Academy of Sciences USA*, doi/10.1073/pnas.1613413114.

Kuper, R.H., Löhr, H., Löning, J. and P. Stehli. 1974. Untersuchungen zur neolithischen Besiedlung der Aldenhovener Platten, IV. *Bonner Jahrbücher* 174: 424-508.

Landes, D.S. 1969. *The Unbound Prometheus.* Cambridge.

Laurent, A. et al., 2019. Ancient pigs reveal a near-complete genomic turnover following their introduction to Europe. *Proceedings of the National Academy of Sciences USA*, doi10.1073/pnas.1901169116.

Mathieson, I. et al. 2015. Genome-wise patterns of selection in 230 ancient Eurasians. *Nature*, doi:10.1038/nature16152.

Menozzi, P., Piazza, A. and L.L. Cavalli-Sforza. 1978. Synthetic mpas of human gene frequencies in Europeans. *Science* 210: 786-92.

Munro, N.D. and M.C. Stiner. 2015. Zooarchaeological Evidence of early Neolithic colonizatioj at Franchthi Cave (Peloponnese, Greece). *Current Anthropology* 56 (4): 596-603.

Olalde, I. et al. 2019. The genomic history of the Iberian Peninsula over the past 8000 years. *Science* 363: 12330-34.

Pinhasi, R., Fort, J. and A.J. Ammerman. 2005. Tracing the origin and spread of agriculture in Europe. *PLoS Biology* 3 (12), e410: 1-9.

Price T.D. and O. Bar-Yosef (eds.). 2011. *The Origins of Agriculture: New Data, New Ideas.* The Wenner-Gren Symposium Series, Chicago.

Racimo F, Sikora M, Vander Linden M, Schroeder H, Lalueza-Fox C (2020) Beyond broad strokes sociocultural insights from the study of ancient genomes. *Nature Reviews Genetics.* https://doi.org10.1038/s41576-020-0218-z.

Rivollat, M. et al. 2020. Ancient geome-wide DNA from France highlight the complexity of interactions between Mesolithic hunter-gatherers and Neolithic farmers. *Science Advances* 6 (22). Eaaz5344.

Skellam, J. 1951. Random dispersals in theoretical populations. *Biometrika* 38: 196-218.

Skellam, J. 1973. The formulation and interpretation of mathematical models of diffusionary processes in biological populations. In: M. Barrlett and R. Hiorns (eds.). *The Mathematical Theory and Dynamics of Biological Populations*, 63-85. London.

Sykes, B. 2003. European ancestry: the mitochondrial landscape. In: A.J. Ammerman and P. Biagi (eds.), *The Widening Harvest. The Neolithic Transition in Europe: Looking back, looking forward.* Archaeological Institute of America, Colloquia and Conference Papers 4, 315-26. Boston.

Szécsényi-Nagy, A. et al. 2014. Ancient DNA evidence for a homogeneous maternal gene pool in sixth millennium cal BC Hungary and the Central European LBK. In A. Whittle and P. Bickle (eds.), *Early Farmers: The View from Archaeology and Science.* Proceedings of the British Academy 198, 71-93. Oxford.

Ucko, P.J. and G.W. Dimbleby (eds.). 1969. *The Domestication and Exploitation of Plants and Animals.* London.

Valdiosera, C. et al. 2018. Four millennia of Iberian biomolecular prehistory illustrates the impact of prehistoric colonization at the far end of Eurasia. *Proceedings of the National Academy of Science USA* 115, 13: 3428-3453 [doi/10.1073/pnas.1717762115].

Vigne, J.D, Briois, F. and M. Tengberg (eds.). 2017. Nouvelles donneées sur les débuts du Néolithique a chypre. Séances de la Société Préhistorique Française. Paris.

Vigne, J.D. et al. 2013. The transportation of mammals to Cyprus sheds light on early voyaging and boats in the Mediterranean Sea. In: A.J. Ammerman and T. Davis (eds.), *Island Archaeology and the Origins of Seafaring in the Eastern Mediterranean.* Proceedings of the Wenner Gren Workshop held at Reggio Calabria on October 19-21, 2012. *Eurasian Prehistory* 10: 157-76.

Whittle, A.  1988.  *Problems in Neolithic Archaeology*.  Cambridge.

Whittle, A. and P. Bickle (eds.).  2014.  *Early Farmers:  The View from Archaeology and Science*. Proceedings of the British Academy 198.  Oxford.

Zilhäo, J.  2001.  Radiocarbon evidence for maritime pioneer colonization at the origins of farming in west Mediterranean Europe.  *Proceedings of the National Academy of Sciences USA* 98: 14180-14185.

Zilhäo, J.  2011.  Time is on my side.  In A. Hadjikoumis, E. Robinson and S. Viner (eds.), *The Dynamics of Neolithisation in Europe.  Studies in honour of Andrew Sherratt*, Oxbow Books, 46-65. Oxford.

Zvelebil, M. and K.V.  Zvelebil.  1988.  Agricultural transition and Indo-European disperals.  *Antiquity* 62: 574-83.